\documentclass[aps,preprintnumbers,nofootinbib,superscriptaddress]{revtex4}

\usepackage{graphicx}
\DeclareGraphicsExtensions{.pdf}

\usepackage{amsfonts,amssymb,amsmath}
\usepackage{amsthm}
\usepackage{color} 

\newcommand{\comment}[1]{}
\newcommand{\ket}[1]{\left |  #1 \right\rangle}
\newcommand{\bra}[1]{\left \langle #1  \right |}

\theoremstyle{plain}

\theoremstyle{definition}

\begin{document}

\title{Deterministic Relativistic Quantum Bit Commitment}

\author{Emily \surname{Adlam}} \affiliation{Centre for Quantum
  Information and Foundations, DAMTP, Centre for Mathematical
  Sciences, University of Cambridge, Wilberforce Road, Cambridge, CB3
  0WA, U.K.}  \author{Adrian \surname{Kent}} \affiliation{Centre for
  Quantum Information and Foundations, DAMTP, Centre for Mathematical
  Sciences, University of Cambridge, Wilberforce Road, Cambridge, CB3
  0WA, U.K.}  \affiliation{Perimeter Institute for Theoretical
  Physics, 31 Caroline Street North, Waterloo, ON N2L 2Y5, Canada.}

\date{April 2015}

\begin{abstract}
We describe new unconditionally secure bit commitment schemes whose
security is 
based on Minkowski causality and the monogamy of quantum entanglement. 
We first describe an ideal scheme that 
is purely deterministic, in the sense that neither party needs
to generate any secret randomness at any stage.
We also describe a variant that
allows the committer to proceed deterministically,
requires only local randomness generation from the
receiver, and allows the commitment to be verified in the
neighbourhood of the unveiling point.  
We show that these schemes still offer 
near-perfect security in the presence of losses and errors,
which can be made perfect if the committer uses an extra single
random secret bit.  
We discuss scenarios where these advantages are significant. 
\end{abstract}

\maketitle

{\bf Introduction} \qquad Relativistic quantum cryptography exploits
the combined power of Minkowski causality and quantum information
theory to control information in order to implement cryptographic
tasks.  A variety of interesting tasks
(e.g. \cite{colbeckkent,bcsummoning,otsummoning,malaney,buhrmanetal,kms,kenttaggingcrypto})
are now known to be achievable, either with unconditional security or
with security significantly enhanced relative to classical protocols.
There has also been progress in characterising fundamental constraints
imposed on quantum information tasks by Minkowski causality
\cite{nosummoning,qtasks,haydenmay}.

The first significant application of relativistic cryptography
was to bit commitment
\cite{kentrel,kentrelfinite,bcsummoning,bcmeasurement}, 
a basic cryptographic primitive 
which has many applications 
and which cannot be implemented securely by using quantum
information alone
\cite{mayersprl,mayersone,lochauprl,lochau,mkp,darianoetal}.  
Several classical and quantum relativistic 
bit commitment protocols have now been proven secure
\cite{kentrelfinite,bcsummoning,bcmeasurement,crokekent,qtasks,kthw,lunghietal}.
The feasibility of secure relativistic quantum
bit commitment has also been demonstrated 
experimentally \cite{lunghietal,liuetal}. 
The feasibility of classical relativistic bit commitment
has also been investigated \cite{kentrelfinite,practicalrbc}
with a view to near term implementation \cite{practicalrbc}.  

Nonetheless, the full range of possibilities
for relativistic quantum bit commitment 
protocols has not yet been systematically explored, 
nor are all the possible tradeoffs between 
security advantages and requirements well understood.  
We are motivated to address these questions both because
they are practically relevant and because the answers
illuminate the general properties of relativistic 
quantum information and its relationship to cryptography.  

Existing relativistic classical and quantum bit commitment
protocols \cite{kentrelfinite,bcsummoning,bcmeasurement} require
at least one party to locally generate and then securely
store and/or distribute secret
classical random strings.   While this is a reasonable capability
to assume in many cryptographic contexts, it may not always be 
practical.  For example, if protocols are being implemented 
over a network of many sites, it may not necessarily
be desirable to set up random number generators or secure
classical memories at every site.   

One might at first think that quantum protocols cannot have
any advantage here, since if a party can securely and reliably prepare,
distribute and measure entangled quantum states, they can obtain
secure classical random strings from those states as and when
required.   In many scenarios this argument may indeed apply.
However, quantum information has security advantages 
compared to classical information, particularly when one
considers a protocol as part of a larger cryptographic exchange.  
For example, if a party is concerned that there has been a security breach at 
one of their sites, they can 
check whether a distributed quantum state remains in the
correct form, whereas they cannot tell for sure whether a purportedly
secret distributed classical random string has been read at some
location by an adversary.  

These points, alongside interest in understanding better theoretically
the relationship between relativistic quantum information and
cryptography, motivate us to consider relativistic quantum
bit commitment protocols that require less secret classical
randomness, or even none.      
We describe here two entanglement-based relativistic
bit commitment protocols that minimize the need for
classical randomness: indeed, one of them, in its ideal
form, requires no randomness at all.   
Their security can be 
understood as a consequence of the monogamy of 
quantum entanglement.   

{\bf Bit commitment} \qquad A bit commitment 
protocol involves two mistrustful parties
who control disjoint secure regions (laboratories) and
exchange information.  The committer, Alice, carries
out actions that commit her to a particular bit
value (or, in the quantum case, a particular superposition
of bit values).  She can later, if she chooses, 
give the receiver, Bob, classical or quantum information that
unveils the committed bit.   Ideally, the protocol 
should rely only on physical principles to 
guarantee to Bob that Alice is committed by her initial
actions, and to Alice that Bob can learn no information
about the committed bit unless and until she unveils.

When considering relativistic bit commitment protocols, 
these definitions need to be framed more carefully \cite{qtasks}. 
In such protocols, both Alice and Bob are 
represented by networks of collaborating
agents distributed appropriately in space-time.   
All of Alice's agents are assumed to be acting 
with perfect trust in one another.   However, at
any given time (in some fixed reference frame), they 
do not necessarily all have the same information, both because
they are separated in space and because quantum information 
cannot be broadcast.
The same applies to Bob's agents.  

In standard relativistic bit commitment protocols, 
the commitment is carried out by 
one of Alice's agents. 
In an idealized model, this agent acts at a single point in 
space-time; more realistically she acts within a spatially small 
secure laboratory during a small time interval. 
The unveiling may be carried out by any number of
{Alice's} agents, 
possibly including the committing agent.   
In principle a protocol could require
agents to follow any specified causal paths in space-time.
However, we usually assume there is a natural inertial frame 
with respect to which they are all stationary, so that they are
located at fixed points in space (or within fixed
small laboratories) throughout the protocol. 
Since we allow arbitrary numbers of agents, this loses
no generality, so long as we assume that Alice's agents have
secure classical and quantum communication channels. 
(Note, however, that this last assumption may not always be 
justified; if not, the possibility of mobile agents 
should be kept in mind.) 

{\bf Security definitions} \qquad 
One needs to be careful about what, precisely,
a bit commitment protocol is intended to guarantee
in relativistic scenarios.   Specifically, one needs to be 
clear which agent (or combination of agents) is (are)
committed at which point(s).  We follow the physically
motivated definition 
first set out in
Ref. \cite{qtasks}, which requires that a bit commitment
should guarantee that 
the committed data was available to and input by 
Alice's committing agent $A_c$ at the space-time point where
the commitment occurs.   This definition allows for the
possibility of $A_c$ inputting a quantum superposition of
the values $0$ and $1$.   However, it excludes protocols
in which the unveiling agents could influence the value
of the unveiled bit by using correlated information that 
they acquired 
independently 
of $A_0$ \cite{qtasks}.\footnote{Following Ref. \cite{qtasks},
another discussion of security definitions from a somewhat
different perspective was given in Ref. \cite{kthw}.}

Let the agents involved in the unveiling be $A_i$ ($i=0,1, \ldots$).
Let $p_0 (S)$ and $p_1 (S)$ be the probabilities that, by following
some collective strategy $S$, they
persuade Bob that, according to the rules of the protocol,
they have validly unveiled $0$ or $1$ respectively.

We say a relativistic quantum bit commitment
protocol is unconditionally secure against Alice if, 
given any commitment actions by $A_c$ that Bob will
accept as valid, and any strategies $S$ and $S'$ by the
unveiling agents $A_i$ that are allowed by quantum theory
and special relativity, we have $p_0 (S) + p_1 (S') < 1 + \epsilon (N)$,
where $N$ is a variable security parameter of the protocol
and $\epsilon (N ) \rightarrow 0$ {as} $N \rightarrow \infty$.  

In the protocols we consider below, there are two unveiling agents
$A_0$ and $A_1$, whose actions are spacelike separated from each
other and from those of $A_c$. 
The probability of a successful unveiling of bit value $i$ depends
only on the actions of agent $A_i$. 
A collective strategy $S$ may be fixed by Alice before the
protocol, or Alice's agents responsible for unveiling $0$ and $1$ may
independently choose their strategies after the commitment time,
possibly conditioned on events in the past lightcone of their
verification point but not of the commitment point. We subsume the
latter possibility under the former by allowing any strategy $S$ to
include steps in which agents make strategic choices with
probabilities conditional on certain external events, with those
events themselves now explicitly included in the description of 
strategy $S$. Any strategy whereupon the conditional
probabilities for these choices are nontrivial may be written as a
convex combination of deterministic strategies, so no probabilistic
strategy can have greater success probability than the most successful
deterministic strategy.  

For protocols of the type we consider
we can thus simplify the above definition: such a protocol provides unconditional
security against Alice if any only if for any collective strategy $S$ which is
possible according to quantum theory and special relativity, $p_0 (S) + p_1 (S)
< 1 + \epsilon (N)$ and $\epsilon (N ) \rightarrow 0$ as $N
\rightarrow \infty$, where $N$ is a variable security parameter of the
protocol, and $p_0 (S)$ and $p_1 (S)$ are the probabilities that, by
following strategy $S$, Alice and her agents persuade Bob that they
have validly unveiled $0$ or $1$ respectively according to the rules
of the protocol. 

We say a relativistic bit commitment protocol 
is unconditionally secure against Bob if, whatever
strategy Bob's agents follow, if Alice's agents choose not to 
unveil, then the probability of any of Bob's agents correctly
guessing the committed bit at any point in space-time
is bounded by $ 1/2 + \epsilon' (N)$, where  
$\epsilon' (N ) \rightarrow 0$ {as} $N \rightarrow \infty$.  
It follows from this definition, by the no-signalling principle,
that when Alice does choose to unveil,
Bob cannot guess Alice’s commitment anywhere that does not lie in the
future lightcone of the unveiling points.

In the protocols we consider below, Alice has one committing agent,
$A_c$, and two unveiling agents, $A_0$ and $A_1$, who can 
unveil a valid commitment to $b=0$ and $1$
respectively.  
An additional security criterion may be required for such protocols:
that if $A_c$ does not make a valid 
commitment to bit value $b$, $A_b$ follows the unveiling
protocol and $A_{\bar{b}}$ does not, then Bob's agents, at any point
in space-time, should gain no information about whether $A_c$
committed to bit value $\bar{b}$ or declined to make a valid
commitment.  As we explain below, with simple modifications,
our protocols also satisfy this criterion. 

{\bf Relation of commitment and unveiling points} \qquad  
Another issue is what exactly is meant by the unveiling 
taking place ``later'' than the commitment in Minkowski space.
In some quantum relativistic bit commitment protocols \cite{bcsummoning,
bcmeasurement}, the unveiling points are in the lightlike 
causal future of the commitment point.  In the idealized
case in which agents are pointlike and their actions are
instantaneous, these protocols guarantee that the committing
agent was committed at the commitment point, in the
sense given above.  
In such protocols, the statement that the unveilings
are later than the commitment is true independent of the
frame.   We call these {\it lightlike causal} (LC) relativistic
bit commitments.   

We wish here also to consider protocols in which 
the unveiling points are space-like separated from
the commitment point.  The most obviously interesting case
is that in which all unveiling points are later than the commitment
point with respect to some fixed frame $F$.  
We call such protocols {\it fixed frame positive duration} (FFPD) 
relativistic bit commitments.  

Generally, if there is a fixed frame $F'$ in which
all the agents are stationary during the protocol, we will take $F'=F$.   
One motivation for considering this case is that it allows us
to consider sequences of protocols in which the
unveiling points tend towards the future light 
cone, and so to relate LC and FFPD commitments.    
Another is that there are many practical
situations -- such as protocols carried out on
terrestrial computer networks -- in which there
is a generally agreed (approximately) inertial
frame and time coordinate.  In such 
scenarios, commitments are potentially useful 
provided they have a positive duration with respect to this
coordinate.   
A third motivation is the possibility of 
sustaining a bit commitment for several rounds by using
sequences of protocols with space-like
separations, as in the examples of Refs. \cite{kentrel,kentrelfinite}. 
In this case, the geometry can be chosen so that any or all possible final
unveiling points are in the causal future of the commitment
point.   A sequence of {LC and/or} FFPD relativistic
bit commitments can thus produce a {\it timelike causal} (TC) relativistic bit
commitment: that is, a commitment in which all the unveiling points
are in the timelike future of the commitment point.  

As usual in quantum cryptography, we initially present our protocols in an 
idealized form assuming perfect quantum state preparations,
transmissions, measurements and computations.  
However, the protocols are 
tolerant to errors and losses, as we discuss later.

{\bf Space-time and communications} \qquad 
We also make standard idealizations about the background geometry and 
signalling speed.   We suppose that space-time is Minkowski
and that Alice and Bob each have agents in secure
laboratories infinitesimally separated from the points $P$, $Q_0$ and
$Q_1$, that signals are sent at precisely light speed, and 
that all information processing is instantaneous.  
Again, these assumptions can be relaxed.  The protocols remain  
secure in realistic implementations with finite separations and 
near light speed communication.  If these corrections are
small, the only significant effect is that 
Bob is guaranteed that Alice's commitment
is binding from some point $P'$ in the near causal future of $P$,
rather than from $P$ itself \cite{bcsummoning}.  
Allowing for small deviations from Minkowski geometry also  
requires small corrections to the geometry when stating the security
guarantees, but does not essentially affect 
security beyond that \cite{kentrelfinite}.  

{\bf Geometry} \qquad Alice and Bob agree on a space-time point $P$,
an inertial set of coordinates $(x,y,z,t)$ for Minkowski space, with
$P$ as the origin.  We focus here on the simplest case in which there
are two possible unveiling points $Q_0$ and $Q_1 $, both space-like
separated from $P$: the protocols straightforwardly extend
{to versions with $N$ unveiling points committing
  ${\rm log} (N)$ bits.}  
Alice and Bob each have agents, who
during the protocol are separated in secure laboratories, adjacent to
each of the points $P$, $Q_0$, $Q_1$.  To simplify for the moment, we
take the distances from these labs to the relevant points as
negligible.  Although it is not necessary for much of our discussion,
we assume that $Q_0$ and $Q_1$ have positive time coordinates in the
given frame, so as to define FFPD relativistic bit commitments.  Let
the agents adjacent to $P$ be $A_c$ and $B_c$, and those adjacent to
$Q_i$ be $A_i$ and $B_i$.
\vskip 10pt
In the following protocols, for 
definiteness, we describe a procedure in which Alice and her agents
exchange qubits by secure physical transportation in the preparation
phase.  However, they may alternatively employ teleportation or a secure
quantum channel without significantly altering the protocols'
security. 
Likewise Bob and his agents may exchange qubits by any
secure means. Bob may also arrange to combine his qubits at a
variety of locations, depending on where he wishes to verify the
unveiled bit. 
\vskip 10pt 
\underline{{\bf {ETBC:}  
Simple Entanglement transfer protocol}} 
 
{\bf Preparation 1.} \qquad $A_c$ 
prepares a total of $2N$ Bell pairs in the state $\Psi^-$;
  let the qubits in the first $N$ pairs be $( W_{0P}^j, W_{0Q}^j )$
  and the second $N$ pairs $(W_{1P}^j, W_{1Q}^j)$, where $j \in [1,
  N]$.  She retains the qubits $W_{iP}^j$, gives the qubits $W_{0Q}^j$
  to $A_0$ and {gives} the qubits $W_{1Q}^j$ to $A_1$.

{\bf 2.} $A_0$ and $A_1$ travel to locations adjacent 
to the spatial coordinates of 
$Q_{0}$ and $Q_{1}$.  We assume that $A_c$, $A_0$ and $A_1$
  have secure laboratories that protect their qubits, 
  so Bob cannot interfere with them in any way 
  after the initial preparation.  In particular, $A_0$ and $A_1$
  travel within secure laboratories.    

{\bf Commitment} At the designated commitment point $P$, $A_c$
gives $B_c$ a set of $N$ labelled
qubits $Q_a^j$.  If she wishes to commit to bit value
$i$, these are the qubits $W_{iP}^j$, for $j \in [1, N]$, 
labelled in sequence.   
 
{\bf Unveiling}   {If the agent $A_i$ believes Alice wishes to unveil, 
  she} gives the labelled qubits
  $W_{iQ}^j$ to Bob's agent $B_i$.  $A_c$ (and/or, if preferred, 
one or both of the $A_i$) also sends to Bob's neighbouring agent
a classical message stating the bit value $b$.  
{(Note that in principle the agents $A_c$, $A_0$ and $A_1$ may make
these decisions independently.   To coordinate them and ensure that
all or none unveil, Alice needs to give them instructions in
advance.  These instructions could depend on separate events in the
past light cones of their unveiling decision points, if Alice knows
these events will be correlated.)}  

{\bf Verification} Once at least one of Bob's agents knows the
  claimed bit value $b$, they securely transmit to one agent (for
  example $B_c$ or $B_b$) all the qubits given to $B_c$ and to $B_b$.
  The receiving agent then carries out projective measurements in the
  Bell basis on the qubits $( Q_a^j , W_{bQ}^j )$ for each $j \in
  [1, N]$.  If they get outcomes corresponding to the Bell state
  $\Psi^-$ for all $j$, Bob accepts that Alice made a valid commitment
  to bit value $b$.   (As noted above, this verification step can
be carried out at a location of Bob's choice: for example, it could
be made by an agent half-way between $B_c$ and $B_b$.) 

{\bf Security against Alice} \qquad We prove security against Alice assuming the
validity of quantum mechanics and assuming that Bob's measuring
devices are reliable.  (Neither this protocol nor the variation
considered below gives Bob device
independent security or security against adversaries who can
exploit hypothetical post-quantum non-signalling theories.)  

Write the Hilbert spaces for the $N$ qubits held by  $B_0$, $B_1$
and $B_c$ 
as $H_0$, $H_1$ and $H_2$ respectively, and write $H = H_1 \otimes H_2
\otimes H_0$.  
Bob tests {for} a
{purported} commitment
{to} zero by a measurement
defined by the projection  
$$
P_0 = \otimes_{j=1}^N ( I_1^j \otimes \ket{\Psi_-}_{20}^j
\bra{\Psi_-}_{20}^j ) \, .  
$$ 
Bob tests {for} a
{purported} commitment
{to} one by a measurement
defined by the projection  
$$
P_1 = \otimes_{j=1}^N ( \ket{\Psi_-}_{12}^j
\bra{\Psi_-}_{12}^j \otimes I_0^j ) \, .  
$$ 
Here $I_k^j$ is the identity operator on the $j$-th qubit in $H_k$
and $\ket{\Psi_-}_{kl}^j$ is a Bell state of the $j$-th qubits in 
$H_k \otimes H_l$. 
The operator $Q = P_0 P_1$ can be written as $Q  = \otimes_{j=1}^N
Q_j$, where $Q_j$ acts on the triple
of $j$-th qubits from each Hilbert space and has operator norm $ | Q_j | = 1/2$;
hence $Q$ has operator norm $ | Q | = 2^{-N}$.
    
For any state $\ket{\psi}$ defining triples of $N$ qubits that 
Alice might hand over to $B_c$, $B_0$ and $B_1$, we thus have 
\begin{eqnarray} 
| Q \ket{ \psi} |  & = &  | P_0 \ket{\psi}  - P_0 ( 1 - P_1 )
\ket{\psi } | \nonumber \\
& \geq &   | P_0 \ket{\psi} |  -  | P_0 ( 1 - P_1 )
\ket{\psi } | \nonumber \\ 
& \geq &   | P_0 \ket{\psi} |  -  | ( 1 - P_1 )
\ket{\psi } |  \, \nonumber \\ 
& \geq &  {( p_0^{1/2} - ( 1 - p_1 )^{1/2} )  )} \, \nonumber
\end{eqnarray}
where $p_0$ and $p_1 $ are the respective probabilities of successfully
persuading Bob that $0$ and $1$ was unveiled using the state
$\ket{\psi}$. 

This gives that 
$p_0 + p_1 \leq 1 + 2^{-N+1} + 2^{-2N}$.  
As this holds for any possible state $\ket{\psi}$, it implies 
security (in the standard sense \cite{kentrelfinite,bcsummoning,bcmeasurement} 
for a relativistic quantum bit
commitment) with security parameter $N$.

{\bf Security against Bob} \qquad At commitment, Bob receives
a set of $N$ qubits entangled with another $N$ qubits not in his
possession.   They have the same reduced state (a uniform mixture)
regardless of the committed bit.   He thus cannot obtain any
information about the bit before unveiling.    

\vskip 10pt 
\underline{{\bf { ETRBC:}  \bf Entanglement transfer
protocol with randomisation}}  \qquad In this variation, Alice 
follows the protocol above, but now 
$B_c$ randomly selects half the qubits given to him to send securely
to $B_0$, sending the other half to $B_1$.   This allows 
both $B_0$ and $B_1$ to directly test the bit value as
soon as they receive these qubits.    
 
{\bf Preparation 1.}  $A_c$ prepares $2N$ Bell pairs, $( W_{0P}^j, W_{0Q}^j )$ and
  $(W_{1P}^j, W_{1Q}^j)$ with $j \in [1, N]$, in the state
  $\Psi^-$. She gives the qubits $W_{0Q}^j$ to $A_0$ and the
  qubits $W_{1Q}^j$ to $A_1$.   We take $N$ even for simplicity.
(The protocol can easily be varied to also allow for odd $N$.) 

{\bf 2.} $A_0$ and $A_1$ travel to locations adjacent 
to the spatial coordinates of 
$Q_{0}$ and $Q_{1}$.  We assume that $A_c$, $A_0$ and $A_1$
  have secure laboratories that protect their qubits, 
  so Bob cannot interfere with them in any way 
  after the initial preparation.  In particular, $A_0$ and $A_1$
  travel within secure laboratories.    

{\bf Commitment} At the designated commitment point $P$, $A_c$
gives $B_c$ a set of N labelled
qubits $Q_a^j$. In order to commit to bit value
$0$, she gives him the qubits $W_{0P}^j$; in order to commit to bit
value $1$, she gives him the qubits $W_{1P}^j$.

{\bf Distribution} $B_c$ sends a randomly selected  size $N/2$ subset $J_0$
of his received qubits to $B_0$ and the remaining subset, $J_1$, to
$B_1$.  All qubits are sent with the corresponding labels $j$. 
 
{\bf Unveiling} 
 {If the agent $A_i$ believes Alice wishes to unveil, 
  she} gives the labelled qubits
  $W_{iQ}^j$ to Bob's agent $B_i$.  ($A_c$ and/or either or both of
  the $A_i$ may also send to Bob's neighbouring agent
a classical message stating the bit value $b$ if they wish, although
it is not necessary in this protocol.   In any case, as in the
previous protocol, some advance instructions from Alice are needed to
ensure any unveiling decisions are coordinated.) 

{ \bf Verification} Once he has received the qubits sent by $B_c$,
  $B_i$ carries out projective measurements in the Bell basis
  on the qubits $( Q_a^j , W_{iQ}^j )$ for each $j \in J_i$. 
  If $B_i$ gets outcomes corresponding to 
  the Bell state $\Psi^-$ for all $j \in J_i$ he accepts that 
  Alice made a valid commitment to bit value $i$.

{\bf Security against Alice}  \qquad Again, we prove security against Alice assuming the
validity of quantum mechanics and assuming that Bob's measuring
devices are reliable.  

Write the Hilbert spaces for the $N$ qubits held by  $B_0$, $B_1$ and $B_c$
as  $H_0$, $H_1$ and $H_2$ respectively, and write $H = H_1 \otimes H_2
\otimes H_0$.  
$B_0$ tests for a commitment of zero by a measurement
defined by the projection  
$$
P^{J_0}_0 = \otimes_{j \in J_0} ( I_1^j \otimes \ket{\Psi_-}_{20}^j
\bra{\Psi_-}_{20}^j ) \, .  
$$ 
$B_1$ tests for a commitment of one by a measurement
defined by the projection  
$$
P^{J_1}_1 = \otimes_{j \in J_1} ( \ket{\Psi_-}_{12}^j 
\bra{\Psi_-}_{12}^j \otimes I_0^j ) \, .  
$$ 
Suppose that Alice prepares a state $\ket{ \psi}$ such that the
probability of passing the test for zero is $p \geq p_0$. 
Then there must be at least one subset $J_0$ for which 
this probability is at least $ p_0$, i.e. for which
$$
p^{J_0}_0 = \bra{\psi} P^{J_0}_0 \ket{\psi} \geq p_0 \, .
$$

Consider any subset $J'_0$ such that $ J_0 \cap J'_0 \leq
{N/3}$. 

By a similar argument to that above, we obtain 
$$
| P^{J_0}_0 P^{J'_1}_1 | \leq {2^{-N/6}} \, .
$$
and 
$$ p^{J'_1}_1 \leq 1 +
{2^{-N/6+1}} + 2^{-N/3} - p^{J_0}_0 \leq 1 - p_0 + {2^{-N/6+1}} + 2^{-N/3} \, . $$

Now the proportion of subsets $J'_0 $ with
 $ J_0 \cap J'_0 > {N/3} $ falls off exponentially with $N$:
to leading order it is bounded by ${(N/6) ( 2^{-10/6}
  3 )^N}$. 
Hence the overall probability of bit value one being accepted, $p_1$,
is bounded by 
$p_1 \leq 1 - p_0 + {2^{-N/6+1} + 2^{-N/3} + O ( N/6 (2^{-10/6} 3
  )^N )}$,
again giving security with security parameter $N$.  

{\bf Security against Bob} \qquad As before, at commitment, Bob receives
a set of $N$ qubits entangled with another $N$ qubits not in his
possession.   They have the same reduced state (a uniform mixture)
regardless of the committed bit.   He thus cannot obtain any
information about the bit before unveiling.    

{\bf Errors and Losses} \qquad In any realistic implementation, Alice's
state preparation and Bob's measurements will be imperfect and their
communication channels and storage devices will have some noise and 
losses.    To show that the protocols will be feasible with sufficiently
good, but imperfect, technology we need versions adapted  
to allow for some non-zero level of errors and losses.  

We first assume that Bob follows the protocol and 
measures each purported singlet separately, and that the errors
and losses for each singlet are small and statistically independent.

For protocol ETBC, in this error model, Bob can test for 
a purported commitment of zero, with negligible probability
of getting a false negative result, by checking that he gets positive
answers for a proportion $( 1 - \epsilon )N$ of tests for the singlet 
$\ket{\Psi_- }_{20}$, where $\epsilon > 0$ is small.
The error model implies that the
probability of a state $\ket{\psi}$ passing the test 
is no more than $ | P_0^{\delta} \ket{\psi}  |^2 + \gamma( \delta, N ) $. 

Here $P_0^{\delta} = \sum_{m= ( 1 - \delta )N}^N P^0_m $, 
where $\delta > \epsilon$ is also small, and chosen so that 
$\gamma (\delta , N ) \rightarrow 0 $ as $N \rightarrow \infty$. The operator $P^0_m$ is the 
projection onto the subspace of states spanned 
by states of the form $ \otimes_{i=1}^N \ket{\Psi_i }_{20} 
\ket{   \Phi_i }_1$, where the $\ket{\Psi_i }_{20} $ are Bell states, of 
which precisely $m$ are $\ket{\Psi_- }$, and the $\ket{ \Phi_i }_1$
are arbitrary qubits in $H_1$. 

Bob similarly tests for a purported commitment of one by checking that he gets positive
answers for a proportion $( 1 - \epsilon )N$ of tests for the singlet 
$\ket{\Psi_- }_{12}$. 
The probability of a state $\ket{\psi}$ passing this test 
is (up to negligible quantities) 
no more than $ | P_1^{\delta} \ket{\psi}  |^2  + \gamma( \delta, N )
$, where $P_1^{\delta} = \sum_{m= ( 1 - \delta )N}^N P^1_m $ is defined
similarly.

The operator  $ P_0^{\delta} $ can be written
as a sum of $\sum_{x=0}^{N \delta} C^N_{N-x} 3^x $ terms involving 
one-dimensional projectors onto tensor products of Bell 
states in $H_2 \otimes H_0$, tensored with the identity on $H_1$.   
The operator  $ P_1^{\delta} $ can be written similarly, using
Bell state projections on $H_0 \otimes H_1$.  
The operator $Q^{\delta} = P_0^{\delta} P_1^{\delta} $
can thus be written as a sum of $( \sum_{x=0}^{N \delta} C^N_{N-x} 3^x
)^2 $ rank one operators, each of which has operator norm
no more than $2^{-N + 2 \delta N}$.  
This gives the (weak, but adequate for our purpose) bound 
$ | Q^{\delta} | \leq 2^{-N + 2 \delta N}  3^{2 \delta N} (N \delta +1 )^2
( C^{N}_{N - N \delta} )^2 $, which tends to zero for large $N$ and
fixed small $\delta$.  The security argument then runs as before. 

The security proof for protocol ETRBC similarly extends to
cover small levels of errors and losses under the assumptions
above.

For completeness, we should note another possible security issue.  
If the errors in Alice's singlet state preparations vary over time in
some predictable way, then the reduced density matrices for the 
states handed over to $B_c$ by $A_c$ may also vary predictably. 
Given a deterministic protocol, we have to assume 
that the order in which $A_c$ labels the singlets after producing
them is public information.
$B_c$ might then be able to infer some information about the
committed bit by measuring these states, without waiting to combine
them with states returned by the $A_i$.    

This may not seem a significant practical worry, since in practice 
one might reasonably expect the predictable component of any variation in
Alice's preparation devices to be very small.  Moreover, some 
deterministic strategies could reduce it further.  For example, the
information revealed by a monotonic drift of some parameter over time 
could be greatly reduced by taking the odd time ordered singlets
produced (the $1$st, $3$rd, and so on) to be the first $N$ for the
protocol, and the even ordered to be the second $N$.  
Still, any predictable variation prevents perfect security against
Bob, according to our definition.   This concern can be eliminated
if $A_c$ groups the states into two batches of $N$ singlets by 
some deterministic method, and then decides randomly which batch
is labelled from $1$ to $N$ and which from $N$ to $2N$.   
This requires her to generate and keep secure a single random bit.  

\vskip 10pt 

{\bf Discussion} \qquad 

{\bf Ideal case: no losses or errors} \qquad   
The first protocol has a theoretically interesting advantage over any previous
relativistic bit commitment protocol in that it is deterministic:
neither party needs to make any random choices of classical data 
or quantum states.  It thus satisfies the strongest possible
form of Kerckhoff's cryptographic principle that a cryptographic
system should be secure even if everything about it except the
choice of key is public knowledge: here, neither party even 
needs a secure key.   Generating secure randomness is 
itself a cryptographic problem that requires extra security
assumptions, or trusted secure quantum devices, or both.
Eliminating any need for it requires fewer resources
and removes some potential security issues.     

These advantages come at a price. 
Bob does not know whether Alice will choose to
unveil a commitment to $0$ or to $1$, and the no-summoning theorem
\cite{nosummoning} prevents him from having the qubit $Q_a$
available at spacelike separated points along the different directions
associated with $0$ and $1$, the time between Alice's unveiling and
the earliest time at which Bob can verify her commitment is twice as
long for this variation.  In time-sensitive situations this may be
a disadvantage. 

This is what motivates the second version of our protocol.
It eliminates this potential
drawback by allowing each $B_i$ to 
test whether the bit is $i$ at the earliest possible point, 
as soon as a light signal from $B_c$ reaches them. 
After these points, 
Alice has essentially zero probability of both persuading
$B_0$ that the bit might be $0$ and $B_1$ that the bit 
might be $1$.   The cost of this advantage is that 
$B_c$ needs to be able to generate a classical random string
that is secure, at least in the sense that Alice
cannot predict it in advance.  The string may be generated
immediately after $B_c$ receives his qubits from $A_c$, and
it does not matter if Alice immediately learns the string.  This is still less 
demanding than requiring Bob to generate a secure random quantum state or
sequence of states and keep its classical description 
secure \cite{bcsummoning,bcmeasurement}.  The protocol also
has an advantage over purely classical relativistic protocols \cite{kentrelfinite} 
in that Alice does not need to generate any secure random data.   

{\bf Losses and errors} \qquad  As shown, our protocols can be modified
to tolerate small losses and errors.  The comments above continue 
to apply, with one small but important qualification.   If Alice
wishes to eliminate any information leaking to Bob because of potentially predictable variation in
Alice's state preparation, our strategy needs $A_c$ to generate and keep secure a
single random bit for each committed bit.  
This is a minimal additional security requirement, and needed only to eliminate for what
in practice might often be a negligible leakage of information.
Still, it should be kept in mind when making comparisons.   

{\bf Need for trusted devices} \qquad  
Both protocols require Bob to rely on his devices
to correctly implement projective measurements for
Bell states, up to known small levels of losses and errors.   
The protocols as stated are thus not fully device independent.
It also follows that they rely for their security on the validity of quantum theory
(not just on the no-signalling principle).  However, the protocols
can be modified to give device
independent versions by replacing verification steps by (for example)
CHSH tests: we will give a detailed discussion elsewhere
\cite{adlamkentdirqbc}.    

{\bf Other comments} \qquad 
Note that, like all technologically unconstrained quantum bit commitment
protocols\cite{kentbccc,kentshort}, our protocols do not 
prevent Alice from committing to a quantum superposition of 
bits.   She can simply input a superposition $\alpha \ket{0} +
\beta \ket{1}$ into a quantum computer programmed to implement
the two relevant quantum measurement interactions for inputs 
$\ket{0}$ and $\ket{1}$ and to send two copies of the quantum
outcome data towards $Q_0$ and $Q_1$, and keep all the data at
the quantum level until (if) she chooses to unveil.   
This gives her no advantage in stand-alone applications of bit
commitment, for example for making a secret prediction: it 
does, however, mean that one cannot assume that in a task
involving bit commitment subprotocols, any unopened
bit commitments necessarily had definite classical bit values,
even if all unveiled bit commitments produced valid classical
unveilings. 

As with the protocols of 
Refs. \cite{kentrelfinite,bcsummoning,bcmeasurement}, the present
protocols can be chained together in sequence, allowing longer term
bit commitments and flexibility in the relation between the commitment
and unveiling sites (in particular, they need not be lightlike
separated).  Full security and efficiency analyses for these chained
protocols remain tasks for future work.

\acknowledgments

This work was partially supported by an FQXi mini-grant and by
Perimeter Institute for Theoretical Physics. Research at Perimeter
Institute is supported by the Government of Canada through Industry
Canada and by the Province of Ontario through the Ministry of Research
and Innovation.


\begin{thebibliography}{10}
\expandafter\ifx\csname url\endcsname\relax
  \def\url#1{\texttt{#1}}\fi
\expandafter\ifx\csname urlprefix\endcsname\relax\def\urlprefix{URL }\fi
\providecommand{\bibinfo}[2]{#2}
\providecommand{\eprint}[2][]{\url{#2}}
\bibitem{broadbenttapp}
A. Broadbent and A. Tapp,
Information-Theoretically Secure Voting Without an Honest
Majority, arxiv:0806.1931.  

\bibitem{mayersprl}
{D.~Mayers, Unconditionally secure quantum bit commitment
is impossible, {\it Phys. Rev. Lett.} {\bf 78} 3414-3417 (1997).}

\bibitem{mayersone}
D. Mayers, Unconditionally secure quantum bit commitment
is impossible, {\it Proceedings of the Fourth Workshop on
 Physics and Computation} (New England Complex System Inst., Boston, 1996),
 p.~226.

\bibitem{lochauprl}
{H.-K.~Lo and H.~Chau, Is quantum bit commitment really
possible?,  {\it Phys. Rev. Lett.} {\bf 78}
3410-3413  (1997).}

\bibitem{lochau}
H.-K.~Lo and H.~Chau, Why quantum bit commitment and
ideal quantum coin tossing are impossible, {\it Proceedings of the
 Fourth Workshop on Physics and Computation} (New England Complex System Inst.,
 Boston, 1996), p.~76.

\bibitem{mkp}
D.~Mayers, A.~Kitaev and J.~Preskill, 
Superselection rules and quantum protocols, {\it Phys. Rev. A} 
{\bf 69} 052326 (2004).
\bibitem{darianoetal}
G.~D'Ariano, D.~Kretschmann, D.~Schlingemann, R.~Werner, 
Reexamination of Quantum Bit Commitment: the Possible and the
Impossible, Phys. Rev. A {\bf 76}, 032328 (2007). 

\bibitem{kentrel}
A.~Kent, Unconditionally secure bit commitment,
{\it Phys. Rev. Lett.} {\bf 83} 1447-1450 (1999).
\bibitem{kentrelfinite}
A.~Kent, Secure Classical Bit Commitment using Fixed Capacity
Communication Channels, 
J. Cryptology  {\bf 18} (2005) 313-335.  

\bibitem{nosummoning}
A.~Kent, 
A No-summoning theorem in Relativistic Quantum Theory,
Quantum Information Processing {\bf 12 (2)} pp 1023-1032 (2013). 
\bibitem{qtasks}
A.~Kent,
Quantum Tasks in Minkowski Space,
Class. Quantum Grav. {\bf 29} (2012) 224013. 
\bibitem{haydenmay}
P.~Hayden and A.~May,
Summoning Information in Spacetime, or Where and When Can a Qubit Be? 
arXiv:1210.0913 
\bibitem{bcsummoning}
A.~Kent,
Unconditionally Secure Bit Commitment with Flying Qudits,
New J. Phys. {\bf 13} 113015 (2011). 
\bibitem{otsummoning}
A.~Kent,
Location-Oblivious Data Transfer with Flying Entangled Qudits,
Phys. Rev. A {\bf 84}, 012328 (2011). 
\bibitem{colbeckkent}
R.~Colbeck and A.~Kent, 
Variable Bias Coin Tossing, 
Phys. Rev. A {\bf 73}, 032320 (2006).  
\bibitem{kenttaggingcrypto}
A.~Kent, 
Quantum Tagging for Tags Containing Secret Classical Data,
Phys. Rev. A {\bf 84}, 022335 (2011). 
\bibitem{malaney}
R.~Malaney, Phys. Rev. A {\bf 81}, 042319 (2010). 
\bibitem{buhrmanetal}
H.~Buhrman et al., arXiv:1009.2490v4 (2011).  
\bibitem{kms}
A.~Kent, W.~Munro and T.~Spiller, 
Quantum Tagging: Authenticating Location via Quantum Information and
Relativistic Signalling Constraints, 
Phys. Rev. A {\bf 84}, 012326 (2011). 
\bibitem{crokekent}
S.~Croke and A.~Kent, Phys. Rev. A {\bf 86}, 052309 (2012). 
\bibitem{kthw} 
J.~Kaniewski et al., IEEE Trans. on Inf. Theory {\bf 59}, 4687-4699
(2013). 
\bibitem{lunghietal} 
T.~Lunghi et al., Phys. Rev. Lett. {\bf 111}, 180504 (2013). 
\bibitem{liuetal}
Y.~Liu et al., Phys. Rev. Lett. {\bf 112}, 010504 (2014). 
\bibitem{practicalrbc}
T.~Lunghi et al., arXiv:1411.4917. 
\bibitem{adlamkentdirqbc}
E.~Adlam and A.~Kent, Device-Independent Relativistic Quantum Bit
Commitment, arxiv:15mm.nnnnn.  
\bibitem{kentbccc}
A.~Kent, Impossibility of unconditionally secure 
commitment of a certified classical bit, {\it Phys. Rev. A} 
{\bf 61} 042301 (2000).
\bibitem{kentshort}
A.~Kent, Why Classical Certification is Impossible in a Quantum World,
Quantum Information Processing, {\bf 11} (2), 493-499 (2012). 
\bibitem{bcmeasurement}
A.~Kent, Unconditionally Secure Bit Commitment by Transmitting
Measurement Outcomes, 
Phys. Rev. Lett. {\bf 109}, 130501 (2012). 

\end{thebibliography}

\end{document}